\documentclass{article}
\usepackage[top=1.5in, bottom=1.5in, left=1.2in, right=1.2in]{geometry}
\usepackage{authblk}
\usepackage{amssymb, amsmath,framed}
\usepackage{graphicx}
\usepackage[round]{natbib}

\usepackage{bm}
\usepackage[colorlinks=true,linkcolor=blue,urlcolor=blue,citecolor=blue,anchorcolor=blue]{hyperref}
\expandafter\ifx\csname package@font\endcsname\relax\else
 \expandafter\expandafter
 \expandafter\usepackage
 \expandafter\expandafter
 \expandafter{\csname package@font\endcsname}
\fi
\def\bq{\begin{equation}}
\def\eq{\end{equation}}
\def\bqy{\begin{eqnarray}}
\def\eqy{\end{eqnarray}}
\def\ga{\gamma}

\def\la{\lambda}

\makeatother

\begin{document}

\title{Characteristics of aquatic biospheres on temperate planets around Sun-like stars and M-dwarfs}

\author{Manasvi Lingam\thanks{Electronic address: \texttt{mlingam@fit.edu}}\,\,}

\affil{Department of Aerospace, Physics and Space Science, Florida Institute of Technology, Melbourne FL 32901, USA}
\affil{Institute for Theory and Computation, Harvard University, Cambridge MA 02138, USA}
\author{Abraham Loeb}
\affil{Institute for Theory and Computation, Harvard University, Cambridge MA 02138, USA}

\date{}

\maketitle

\begin{abstract}
Aquatic biospheres reliant on oxygenic photosynthesis are expected to play an important role on Earth-like planets endowed with large-scale oceans insofar as carbon fixation (i.e., biosynthesis of organic compounds) is concerned. We investigate the properties of aquatic biospheres comprising Earth-like biota for habitable rocky planets orbiting Sun-like stars and late-type M-dwarfs such as TRAPPIST-1. In particular, we estimate how these characteristics evolve with the available flux of photosynthetically active radiation (PAR) and the ambient ocean temperature ($T_W$), the latter of which constitutes a key environmental variable. We show that many salient properties, such as the depth of the photosynthesis zone and the net primary productivity (i.e., the effective rate of carbon fixation), are sensitive to PAR flux and $T_W$ and decline substantially when the former is decreased or the latter is increased. We conclude by exploring the implications of our analysis for exoplanets around Sun-like stars and M-dwarfs.
\end{abstract}

\section{Introduction} \label{SecIntro}
It is a well-known fact that the Earth's environment---its lithosphere, hydrosphere, atmosphere, and biosphere---has transformed greatly over time \citep{lun13,Knoll,KN17,SSC20}, and the same also applies to other terrestrial planets in our Solar system \citep{Eh16,KAC19}. In tandem, there is growing awareness and acceptance of the fact that habitability is a multi-faceted and dynamic concept that depends on a number of variables aside from the existence of liquid water \citep{Dole,Kast12,Co16,SBJ16,Co20,LL21}; the latter criterion has been widely employed to demarcate the limits of the so-called habitable zone and its manifold extensions \citep{Hu59,Dole,KWR93,KRK13,KRS14,Ram18,RAF19,SRO19}. 

One of the most crucial environmental parameters that regulates myriad biological processes, and thus the propensity for planetary habitability, is the ambient temperature \citep{CB87,HS02,Ang09,AC17}. It is not surprising, therefore, that there exists a large corpus of work on the thermal limits of life based on comprehensive experiments on thermophiles \citep{RM01,McK14,AC14,MAB19}. In recent times, numerical models have employed the thermal limits for Earth-like complex life to assess the habitability of exoplanets for such organisms \citep{SVS17,MPV20}, and similar analyses have been undertaken for generic subsurface biospheres as well \citep{MOP13,Mana20}.

Motivated by these facts, we will study how temperature impacts the prospects for aquatic photosynthesis on Earth-analogs around stars of two noteworthy spectral types. By Earth-analogs, we refer hereafter to rocky planets that are sufficiently similar to Earth insofar as \emph{all} their geological, physical, and chemical properties are concerned. Our reasons for choosing to investigate aquatic photosynthesis are twofold. First, the importance of photosynthesis is well-established from the standpoint of physics and biochemistry as stellar radiation is the most plentiful source of thermodynamic disequilibrium \citep{DW10}, and photosynthesis represents the dominant avenue for the biosynthesis on organic compounds on Earth \citep{BPM18}. 

In particular, we will focus on oxygenic photosynthesis because its electron donor (water) is available in plentiful supply, consequently ensuring that this mechanism is not stymied by the access to electron donors \citep{WRF}. Moreover, the advent of oxygenic photosynthesis is known to have profoundly altered Earth's geochemistry and biology \citep{Lan02,Jud17}. We will adopt the conventional range of $\lambda_\mathrm{min} = 400$ nm and $\lambda_\mathrm{max} = 700$ nm for oxygenic photosynthesis \citep[Chapter 1.2]{Bla14}, known as photosynthetically active radiation (PAR). To be precise, oxygenic photosynthesis can operate at wavelengths of $350$-$750$ nm \citep{CB11,NMS18,CAB21}, but the canonical choice of the PAR range delineated above does not alter our subsequent results significantly.

We will not delve into the feasibility of multi-photon schemes that might elevate $\lambda_\mathrm{max}$ to longer wavelengths, because their efficacy has not been adequately established. On the one hand, it is plausible that the upper bound (namely $\lambda_\mathrm{max}$) for PAR could be boosted to wavelengths of $\gtrsim 1000$ nm \citep{WoRa,TRY06,KST07,Man19,Man20}. However, on the other hand, these multi-photon schemes may be more fragile and susceptible to low efficiencies due to side reactions \citep{KST07,Kume19,LiM20}. Moreover, recent numerical modeling based on empirical data indicates that, while photosynthesis in the near-infrared is feasible, oxygenic photosynthesis on M-dwarfs may eventually revert to the conventional PAR range described in the preceding paragraph \citep{GW17,TMT17}.

The second reason why we opt to investigate the prospects for aquatic photosynthesis stems from the basic datum that the oceans contribute nearly half to the overall NPP of modern Earth \citep{FB98}. In fact, Earth was almost exclusively composed of oceans (i.e., virtually devoid of large landmasses) for a certain fraction of its history \citep{IK10,AN12}, implying that aquatic photosynthesis may have played an even more significant role in those periods. A few theoretical models have even proposed that continents only emerged in late-Archean era in the neighborhood of $2.5$ Gya \citep{FCR08,LL21}; this conjecture seems to be compatible with the recent analysis of oxygen-18 isotope data from the Pilbara Craton of Western Australia \citep{JW20}.

Looking beyond Earth, statistical analyses of exoplanets indicate that a substantial fraction of super-Earths are rich in volatiles \citep{Rog15,WL15,ChKi17,ZJS18,JM18}. In particular, some of the Earth-sized planets in the famous TRAPPIST-1 system \citep{GT17} may fall under this category, with the water fraction potentially reaching values as high as $\sim 10\%$ by mass \citep{GDG18,UHD18,DMG18}. The habitability of ocean planets (also called water worlds), which are wholly devoid of continents \citep{Kuc03,LS04}, has been analyzed from multiple standpoints \citep{ACC12,KSR13,CA14,Gold15,NSR17,KF18,RL18,LiMa}. In recent times, increasing attention is being directed toward oceanographic phenomena such as salinity, circulation patterns and nutrient upwelling \citep{HY14,CS16,LL18,YAK19,COJ19,DWK19,OJA,SOK20} on such worlds. However, a detailed treatment of the salient characteristics of aquatic photosynthesis remains missing for the most part.

It is important to recognize that we will deal with aquatic environments, but this does not necessarily imply that all worlds under consideration must be solely composed of oceans. The outline of the paper is as follows. We commence with a description of some of the basic tools needed to facilitate our analysis in Sec. \ref{SSecPrelim}. We proceed thereafter by calculating how the properties of aquatic photosynthesis such as the compensation depth and the net primary productivity (NPP) vary with the PAR flux and ocean temperature in Sec. \ref{SecChar}. Next, we explain the salient model limitations in Sec. \ref{SecModLim}. Subsequently, we delineate the ramifications arising from our modeling for Earth-like exoplanets in Sec. \ref{SecDisc}, and we conclude with a synopsis of our central findings in Sec. \ref{SecConc}. 

\section{Mathematical preliminaries}\label{SSecPrelim}
In order to study the basic characteristics of aquatic photosynthesis and their dependence on the average ocean temperature ($T_W$), we hold all parameters (biological, geological and astrophysical) identical to that of Earth. We consider two different Earth-analogs hereafter: one around a solar twin (Planet G) and the other around a late-type M-dwarf (Planet M) with effective temperatures of $T_\odot = 5780$ K and $T = 2500$ K, respectively. Planet M is taken to be \emph{tidally locked} \citep{Barn17}, and the star that it orbits has a temperature closely resembling that of TRAPPIST-1 \citep{DGT18}. The reason for doing so is that Sun-like stars are considered ``safe'' targets for biosignature searches \citep{KWR93,HA14,LiL18}, whereas the habitability of M-dwarf exoplanets, especially those orbiting active stars, remains subject to many ambiguities \citep{TB07,SKS07,SBJ16,LL19}.

In what follows, we draw upon two major simplifying assumptions. First, we model the star as an idealized black body with an effective temperature of $T$. Second, we account for the attenuation of PAR after the passage through the atmosphere by introducing a fudge factor. While neither of these simplifications are entirely realistic, the global results are known to deviate from more realistic models and data by a factor of only $< 1.5$ for the most part \citep{LL20}.\footnote{In fact, the spatial heterogeneity inherent to oceans are known to introduce local variations that are more than an order of magnitude greater than this factor \citep{BBS05}, owing to which the estimates in \citet{LL20} can be regarded as fairly accurate global values.} The reason for this reasonable degree of accuracy stems from the fact that most of the basic quantities we compute hereafter exhibit a weak (i.e., semi-logarithmic) dependence on the two assumptions outlined above.

As we are dealing with Earth-analogs, the stellar flux at the planet's location is taken to be $S_\oplus \approx 1360$ W/m$^2$. At the substellar point on the planet, the photon flux density ($\mathcal{N}_\mathrm{max}$) at the top of the atmosphere is given by
\begin{equation}\label{SpecRadDef}
   \mathcal{N}_\mathrm{max}(\lambda) \approx n_\lambda \left(\frac{R_\star}{d_\star}\right)^2,
\end{equation}
with $R_\star$ and $d_\star$ constituting the stellar and orbital radius, respectively, whereas $n_\lambda$ is the photon flux density of the star at its surface. The black body brightness $B_\lambda$ is invoked to yield
\begin{equation}
 n_\lambda = \frac{B_\lambda}{(hc/\lambda)} = \frac{2c}{\lambda^4}\left[\exp\left(\frac{h c}{\lambda k_B T}\right)-1\right]^{-1},
\end{equation}
where $\lambda$ is the photon wavelength. As we have assumed the stellar flux is equal to $S_\oplus$ for the Earth-analogs, we can express $d_\star$ as follows:
\begin{equation}
    d_\star = \sqrt{\frac{L_\star}{4\pi S_\oplus}}
\end{equation}
where the stellar luminosity ($L_\star$) is given by $L_\star = 4\pi \sigma R_\star^2 T^4$. After employing this relation in (\ref{SpecRadDef}), we find that $ \mathcal{N}_\mathrm{max}$ transforms into
\begin{equation}\label{SpecRadMax}
    \mathcal{N}_\mathrm{max}(\lambda) \approx \frac{n_\lambda S_\oplus}{\sigma T^4}.
\end{equation}
It is, however, necessary to recognize that $\mathcal{N}_\mathrm{max}$ constitutes an upper bound for the photon flux density at the surface for two reasons. First, this photon flux density is calculated at the zenith, and therefore ignores the fact that a given location will not always correspond to the substellar point. Second, the effects of clouds and atmospheric attenuation are neglected. Hence, a more viable expression for the photon flux density at the planet's surface ($\mathcal{N}_\mathrm{avg}$) is given by
\begin{equation}\label{SpecRadavg}
    \mathcal{N}_\mathrm{avg}(\lambda) \approx \mathcal{N}_\mathrm{max}(\lambda) \cdot f_\mathrm{I} \cdot f_\mathrm{CL},
\end{equation}
with $f_\mathrm{CL}$ embodying the total atmospheric attenuation \citep[Chapter 4.2]{SG06}, and $f_\mathrm{I}$ quantifying the average intensity of light at a given location as a fraction of the intensity at the substellar point. Henceforth, we adopt $f_A \equiv f_\mathrm{I} \cdot f_\mathrm{CL} \approx 0.2$ for reasons elucidated further in \citet{LL20} and to maintain compatibility with Earth's global parameters \citep[Chapter 4.3]{SG06}; altering this fraction by a factor of order unity does not change our results significantly due to the logarithmic dependence alluded to earlier in this section. With this choice of $f_A$, it should be noted that $\mathcal{N}_\mathrm{avg}(\lambda) \approx 0.2 \mathcal{N}_\mathrm{max}(\lambda)$. 

Although the above choice has been motivated in \citet{LL20}, a recapitulation is warranted at this stage. On the one hand, $f_\mathrm{I}$ is higher for M-dwarf exoplanets due to the fact that the tidally locked dayside does not experience nights and is bathed in continual illumination \citep{GW17}. On the other hand, $f_\mathrm{CL}$ is reduced as a consequence of the potentially higher atmospheric absorptivity and increased cloud clover, among other factors \citep{KWR93,YCA13,KWH16}. Therefore, by specifying $f_A$ to be constant (as we did in the previous paragraph), we are effectively already ensuring that the atmospheric attenuation experienced by M-dwarf Earth-analogs is a few times higher than their counterparts around Sun-like stars, in line with prior theoretical predictions.

We have verified that quantities such as the compensation depth, the critical depth, and the net primary productivity (all of which are defined later) decrease by a factor of $\lesssim 2$, \emph{ceteris paribus}, even up to nearly an order of magnitude increase in the degree of atmospheric attenuation. Lastly, we remark that the interplay of all the aforementioned variables is further complicated by the presence of climate feedback mechanisms as well as the atmospheric and surface compositions that may collectively yield different values from one climate model to another, even for the same setup, which makes estimating them challenging \citep{ZSDS,SBJ16,CS16}. Assessing the properties of aquatic photosynthesis is a complicated task, as elucidated in Sec. \ref{SecModLim}, owing to which our goal herein is to primarily focus on understanding how the salient characteristics vary as a function of key \emph{physical} parameters that can be constrained by present-day or forthcoming observations \citep{FAD18}.

Given the photon flux density, denoted by $\mathcal{N}_0(\lambda)$, at the surface, we are in a position to calculate the photon flux $\mathcal{F}$ at a depth $z$ below the surface of the ocean. This quantity is found by convolving $\mathcal{N}_0(\lambda)$ and the vertical attenuation coefficient $K$ in the oceans, thus yielding
\begin{equation}\label{IntRel}
    \mathcal{F}(z) \approx \int_{\lambda_\mathrm{min}}^{\lambda_\mathrm{max}} \mathcal{N}_0(\lambda) \exp\left(-K z\right)\,d\lambda.
\end{equation}
It should be noted that $\mathcal{N}_0(\lambda)$ is equal to $\mathcal{N}_\mathrm{max}$ or $\mathcal{N}_\mathrm{avg}$, depending on what scenario we wish to analyze. Now, let us turn our attention to $K$, which we shall rewrite as $K = K_W + K_I$ \citep[Chapter 9.5]{Kirk11}. The first term ($K_W$) is the attenuation coefficient associated with water whereas $K_I$ accounts for the attenuation stemming from impurities as well as biota. In order to tackle $K_W$, we begin by noting that it has been tabulated as a function of $\lambda$ in many sources \citep{HQ73,SB81,KLC93,LQF99,MGC07}. Based on the data taken from \citet[Table 3]{PF97}, which is consistent with later studies over the PAR range \citep{LWV15}, the following simple exponential fit was employed by \citet{LL20} across the PAR range:
\begin{equation}
    K_{W,22} \approx 1.4 \times 10^{-5}\,\mathrm{m}^{-1}\,\exp\left(\lambda \cdot 1.54 \times 10^7\,\mathrm{m}^{-1}\right),
\end{equation}
although it is essential to recognize that the data had been collected at $22$ $^\circ$C \citep[Table 3]{PF97}. In general, $K_W$ is not only dependent on $\lambda$ but also on $T_W$. The ocean temperature in turn varies with depth, but it only changes by a few K in the zone where the bulk of photosynthesis occurs \citep{Paw13}. Hence, we shall treat $T_W$ as being roughly constant, thereby enabling us to model it as a free parameter in the model. In order to account for the dependence on $T_W$, we employ the linear temperature scaling that has been confirmed by a number of empirical studies \citep{LMQ01,STZ06,RRM14}, thereupon enabling us to write
\begin{equation}
    K_W(T_W,\lambda) = K_{W,22} + \alpha(\lambda) \Delta T_{22},
\end{equation}
where $\alpha(\lambda)$ represents the wavelength-dependent temperature coefficient (units of m$^{-1}$ K$^{-1}$), and $\Delta T_{22} = T_W - 295$ is a measure of the deviation from the standard water temperature of $22$ $^\circ$C employed in calculating $K_{W,22}$.

For the PAR range considered herein, the second term on the right-hand-side of the above expression is always much smaller than the first term provided that $\Delta T_{22}$ is $\mathcal{O}(10)$ K. This condition arises because $\alpha$ is nearly zero across the PAR range ($\lesssim 10^{-3}$ m$^{-1}$ K$^{-1}$), as can be verified by comparing \citet[Figure 5]{RRM14} and \citet[Table 1]{STZ06} with \citet[Table 3]{PF97}. Apart from the temperature dependence, we remark that $K_W$ also exhibits a dependence on the salinity, which is naturally expected to vary from one ocean to another \citep{CS16,OJA}. However, we have implicitly held the salinity fixed to that of the global value of Earth's oceans. More importantly, the salinity dependence is weak across the PAR range, as shown by experimental studies \citep{STZ06,RRM14}. 

\begin{figure*}
$$
\begin{array}{cc}
  \includegraphics[width=7.3cm]{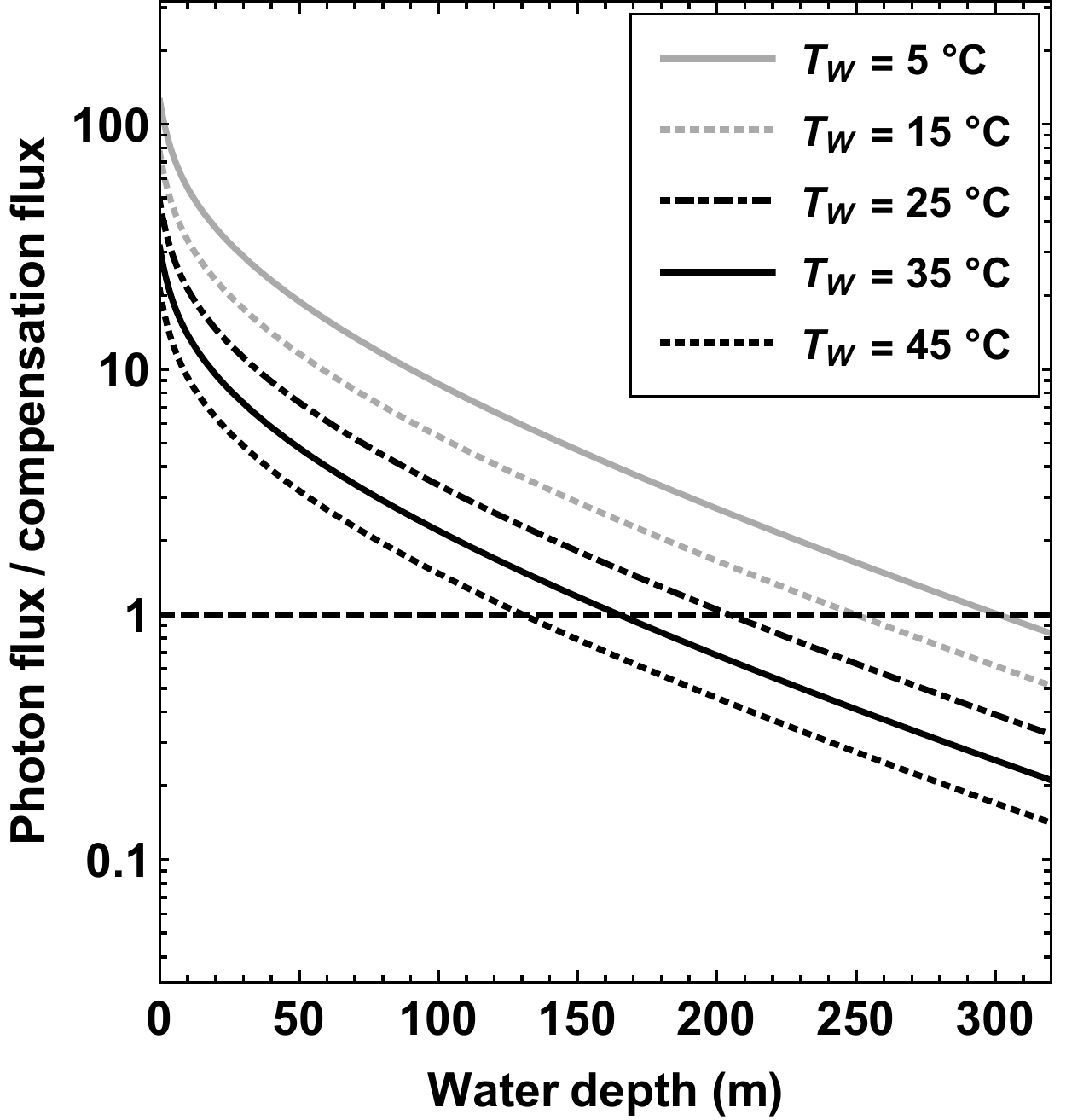} & 
  \includegraphics[width=7.5cm]{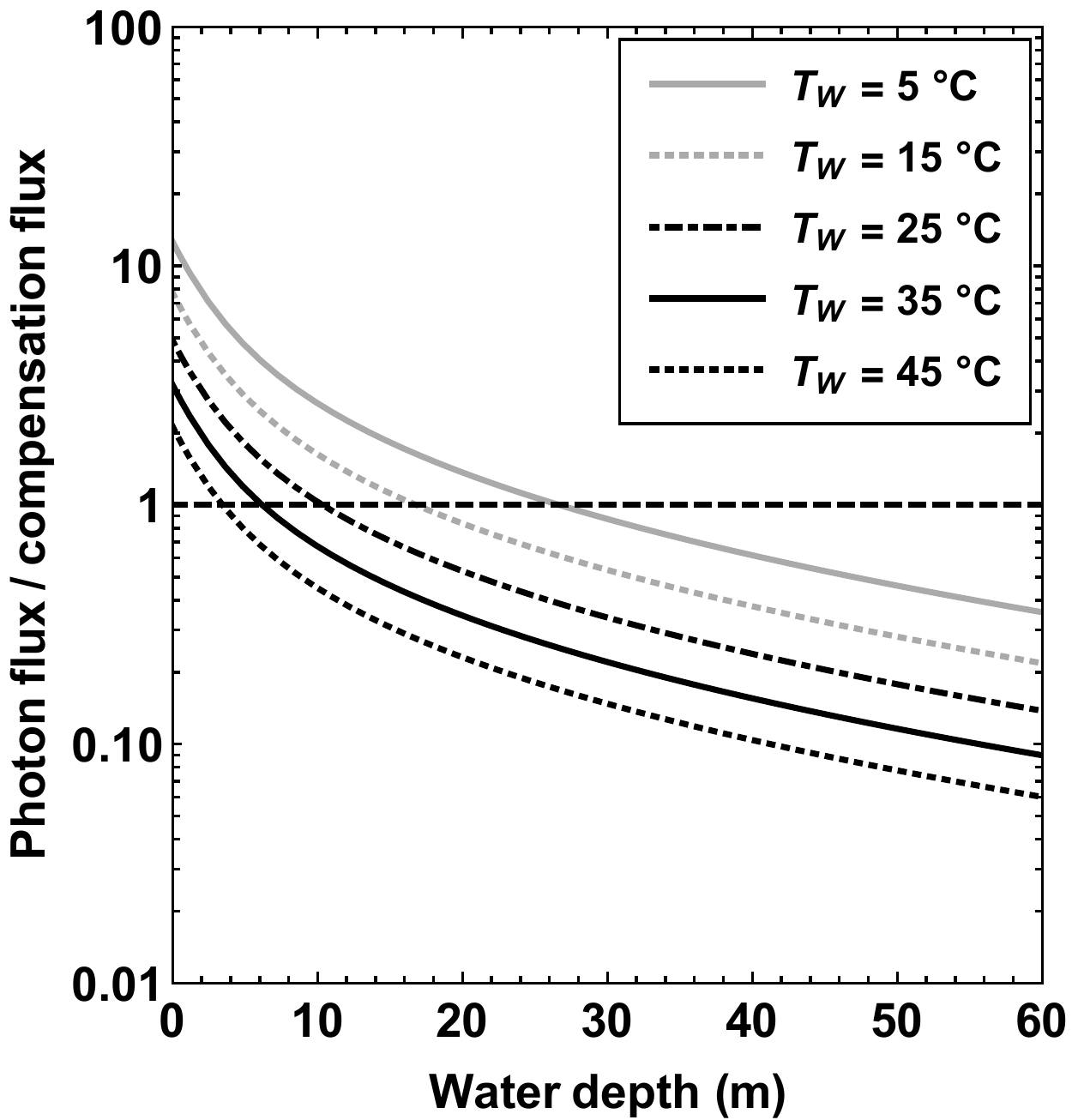}\\
\end{array}
$$
\caption{In both panels, the photon flux in units of the compensation flux is shown as a function of the ocean depth for the idealized case described in Sec. \ref{SSecPrelim}; the compensation flux specifies the photon flux at which net growth of the organism is not feasible. The various curves correspond to different choices of the global ocean temperature, and the intersection points of the various curves with the dashed horizontal line yield the compensation depths. The left and right panels correspond to Planets G \& M, respectively, introduced in Sec. \ref{SSecPrelim}.}
\label{FigCDepG}
\end{figure*}

\begin{figure*}
$$
\begin{array}{cc}
  \includegraphics[width=7.5cm]{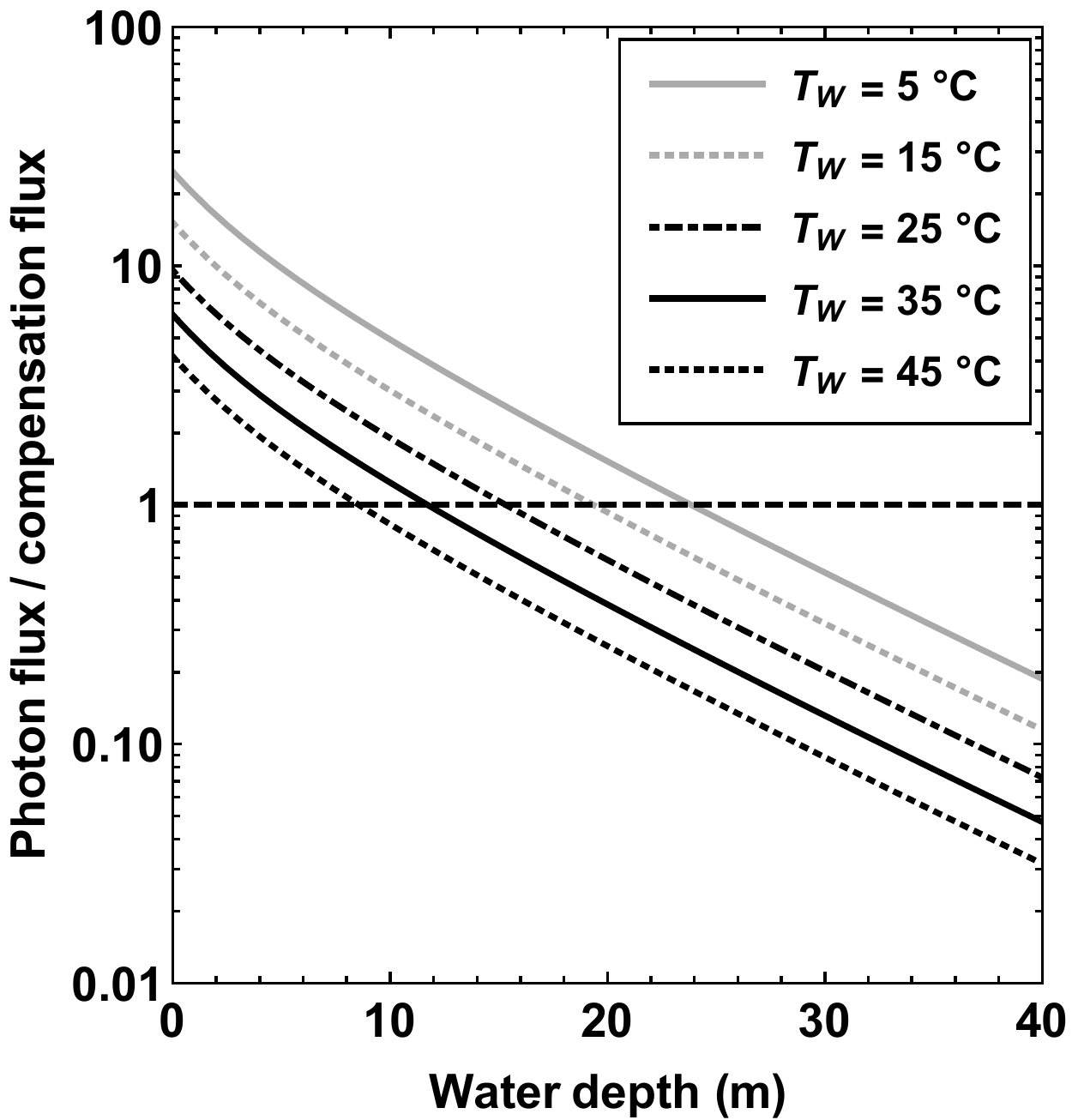} & 
  \includegraphics[width=7.4cm]{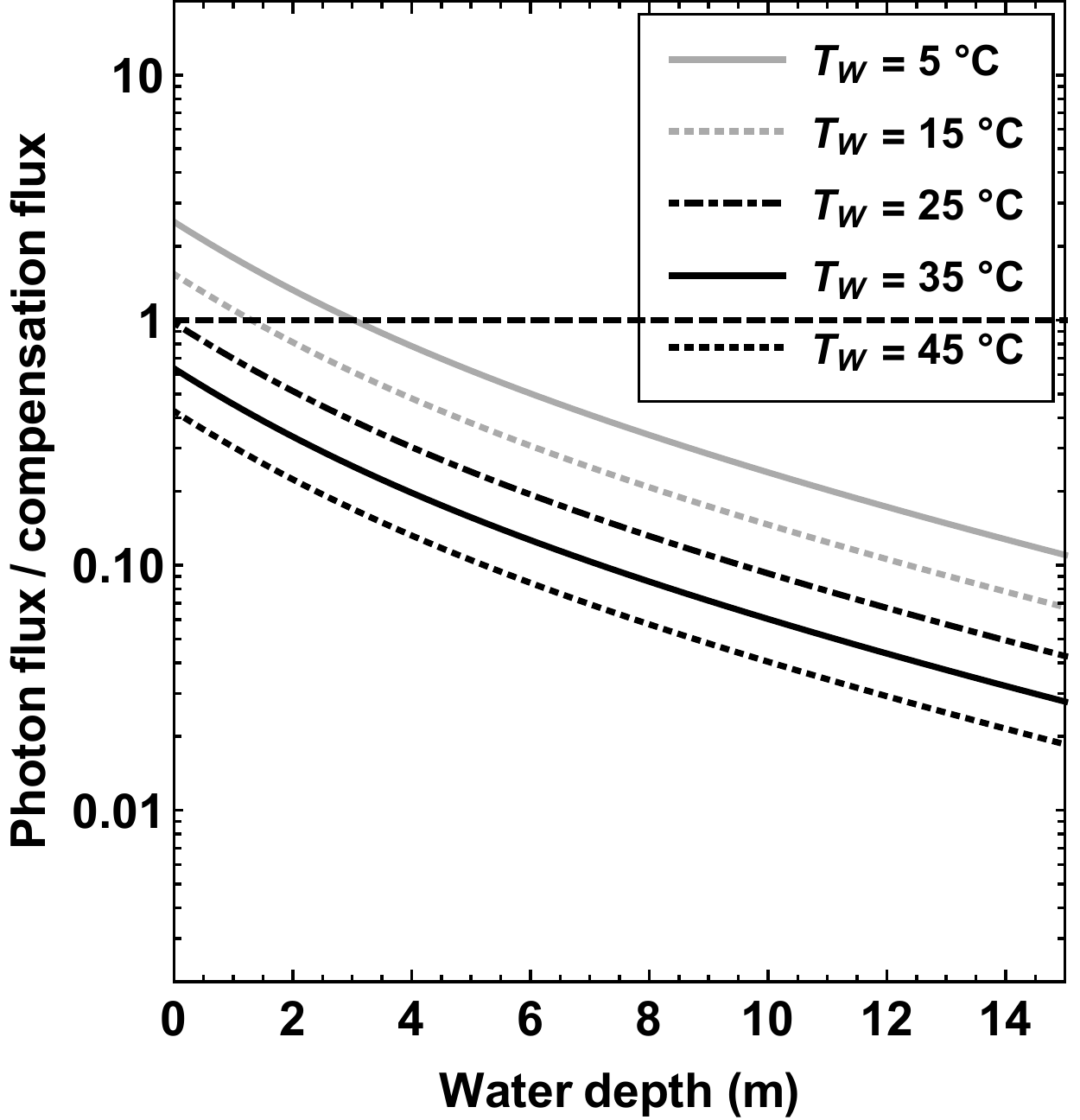}\\
\end{array}
$$
\caption{In both panels, the photon flux in units of the compensation flux is shown as a function of the ocean depth for the global average case described in Sec. \ref{SSecPrelim}; the compensation flux specifies the photon flux at which net growth of the organism is not feasible. The various curves correspond to different choices of the global ocean temperature, and the intersection points of the various curves with the dashed horizontal line yield the compensation depths. The left and right panels correspond to Planets G \& M, respectively, introduced in Sec. \ref{SSecPrelim}; for Planet M, the global average encapsulates the properties on the dayside.}
\label{FigCDepM}
\end{figure*}

Next, we turn our attention to the other attenuation coefficient $K_I$. If one considers the case with pure water, i.e., amounting to $K_I \rightarrow 0$, it follows that $\mathcal{F}$ is maximized for a given depth \emph{ceteris paribus}. In a more realistic setting, however, we shall adopt $K_I \approx 0.08$ m$^{-1}$ to maintain consistency with the typical diffuse attenuation coefficient in the PAR range for Earth's oceans \citep[Chapter 4.2]{SG06}; this choice is also compatible with the coefficients deduced from measurements of clear ocean waters \citep{LDM05,SHG13,SW15}. In actuality, $K_I$ will also be a function of wavelength and temperature, but the exact dependence is dictated by several complex oceanographic and biological (e.g., density of photoautotrophs) factors \citep{MGC07}, owing to which we have opted to work with a constant value. The wavelength variation, in particular, is rather weak because $K_I$ changes by only a factor of $\sim 2$ across the PAR range \citep[Figure 4]{MM01}.

At this stage, it is worth recapitulating the two broad scenarios we shall be considering. The first corresponds to the so-called idealized case where the star is located at the substellar point, and there is no attenuation because of the atmosphere and oceanic impurities. In other words, we employ $\mathcal{N}_0(\lambda) = \mathcal{N}_\mathrm{max}$ and $K = K_W$, and introduce the superscript ``I'' (for idealized). This outcome was studied extensively by \citet{RLR18}, albeit with an exclusive focus on Earth and Proxima b. The second case accounts for time-averaged stellar flux and the existence of biological attenuation. Here, we select $\mathcal{N}_0(\lambda) = \mathcal{N}_\mathrm{avg}$, $K = K_W + K_I$ and the non-zero value of $K_I$ defined in the prior paragraph, and label it using the superscript ``A'', to wit, the ``global average'' case. For each of these two scenarios, we consider two different Earth-analogs (Planets G \& M) delineated at the beginning of this section. For Planet M, however, the ``global average'' refers to the \emph{dayside} parameters, e.g., the variable $T_W$ represents the average temperature on the dayside of the tidally locked M-dwarf exoplanet.

\section{Characteristics of aquatic biospheres}\label{SecChar}
In this section, we examine how certain the salient properties of aquatic biospheres depend on the ocean temperature; in some cases, we investigate the joint dependence on stellar and ocean temperatures. Before embarking on the discussion, we will define the quantities of interest that appear herein; for a historical treatment, we defer to \citet{Mi12,BB18}.

The first concept that we introduce from biological oceanography is the euphotic zone depth: the location where the photon flux becomes $1\%$ of its surface value. As one can see from the definition, it is divorced from biological properties for the most part. The euphotic zone depth is commonly interpreted as a measure of the photosynthesis zone on Earth, but it does not constitute a reliable metric in actuality \citep{Ban04,MLV14}. Before proceeding ahead, we note that the depth of the euphotic zone decreases from $\sim 10$-$100$ m for Earth-analogs orbiting Sun-like stars to $\mathcal{O}(1)$ m for tidally locked late-type M-dwarf exoplanets \citep{RLR18,Kal19,LL20}.

Next, we consider the compensation depth ($\mathcal{Z}_\mathrm{CO}$), which is determined by calculating the location at which $\mathcal{F}(z)$ is equal to the compensation flux ($\mathcal{F}_C$). The latter is roughly defined as the flux at which the rate of growth via photosynthesis becomes equal to the rate of respiration \citep{GG27,MO28}; in other words, at this depth, the net growth rate of the photoautotroph under consideration is equal to zero at $\mathcal{Z}_\mathrm{CO}$. As per the definition, the compensation depth is regulated by $\mathcal{F}(z)$, which in turn, \emph{inter alia}, depends on the parameter $f_\mathrm{I}$ introduced in Sec. \ref{SSecPrelim}. Here, it is important to appreciate that $f_\mathrm{I}$ is \emph{different} for M- and G-type exoplanets \citep{LL20}, due to the fact the dayside of the former receives permanent illumination when tidally locked (amounting to higher $f_\mathrm{I}$ broadly speaking). Although $f_\mathrm{I}$ functions as a fudge factor to an extent, we acknowledge that it does not fully capture the distinct spatiotemporal differences in light distribution, or oceanic properties like nutrient upwelling \citep{LL18,OJA}, associated with tidally locked M-dwarf exoplanets.

The other quantity of interest is the critical depth ($\mathcal{Z}_\mathrm{CR}$), which was elucidated by \citet{GB35} and placed on a quantitative footing by \citet{Ril46} and \citet{Sver53}. It can be envisioned as the integrated (i.e., global) version of the compensation depth. The critical depth is the location where the \emph{vertically integrated} net growth rate becomes zero, i.e., the integrated photosynthetic growth rate is equal to the integrated depletion rate arising from respiration and other factors \citep[Chapter 3]{ML06}. The critical depth is relevant from an observational standpoint because it may regulate the feasibility of phytoplankton blooms \citep{FR07}, which have been posited as an example of temporal biosignatures \citep{LL18,SKP18}. If the ocean mixed layer depth is greater than the critical layer depth,\footnote{As the name indicates, the mixed layer refers to the region of the ocean that is characterized by nearly uniform characteristics (e.g., temperature and salinity), and is governed by the vertical potential density gradient \citep{Kirk11,Mid19}.} the initiation of phytoplankton blooms is rendered unlikely, and vice-versa \citep[pg. 94]{ML06}. Although the critical depth concept remains influential and useful to this day \citep{NS91,OIE,SDY02,Chi11,Kirk11,FMA14,SJB15}, it has been subjected to some criticism \citep{SP90,Behr10,BB18}.

Thus, broadly speaking, the compensation depth and the critical depth represent important concepts inasmuch as exo-oceanography is concerned because they enable us to gauge the depths at which photosynthetic organisms can exist and/or give rise to tangible biosignatures \citep{SJF42,Sver53}. We refer to \citet[Figure 9.5]{FR07} for a schematic overview of these two quantities along with the euphotic zone depth.

\subsection{Compensation depth}\label{SSecCoD}
The key point worth appreciating when it comes to the compensation depth is that the compensation flux ($\mathcal{F}_C$) is \emph{not} constant even for a given organism because it is intrinsically temperature-dependent. Thus, our chief objective is to find a suitable function that will adequately describe the behavior of $\mathcal{F}_C$ with respect to $T_W$. 

In the classical model for the compensation flux, it is proportional to $\Gamma_R/\Gamma_P$---see \citet{Ril46}, \citet{Sver53}, \citet[Equation 2]{SDY02} and \citet[Chapter 3]{ML06}---where $\Gamma_R$ and $\Gamma_P$ signify the rates of respiration and photosynthesis, respectively. Thus, if we know how these rates vary with temperature, one can duly formulate the expression for $\mathcal{F}_C$. The temperature dependence of these rates is subject to uncertainty and many different fitting functions have been considered. However, both the well-known metabolic theory of ecology \citep{GBW01,BGA04,DPS11,BCC15,AC17} and recent analyses of empirical data from Earth's oceans \citep[Figure 3.3]{Kir18} predict that these rates are fairly well described by the classic Arrhenius equation.

Hence, by utilizing the respective activation energies for these two processes \citep[Section 3]{RD12}, we end up with
\begin{equation}\label{GamRat}
    \frac{\Gamma_R}{\Gamma_P} \propto \exp\left(-\frac{\Delta E}{k_B T_W}\right),
\end{equation}
where $\Delta E \approx 0.34$ eV constitutes the ``net'' activation energy, i.e., the difference between the corresponding activation energies \citep{YDC12}. An important point worth noting is that the above ansatz for $\Gamma_R/\Gamma_P$ is monotonically increasing with temperature. It is very unlikely that this behavior would be obeyed \emph{ad infinitum} because the Arrhenius equation breaks down beyond a certain temperature \citep{King09,Ang09,Sch15}. The issue, however, is that the optimum temperature, after which the trend reverses, is species-dependent \citep{AC17,CMB18}, and is modulated to a substantial degree by the environment(s) of the putative organisms. We will restrict ourselves to $273 < T_W < 323$ K, as this interval roughly overlaps with the temperature range of $280 < T_W < 322$ K studied in \citet[pg. 724]{BJ20}. In that study, diverse marine phytoplankton were shown to obey (\ref{GamRat}) for a broad thermal range.

By utilizing the above relationships, the temperature dependence of $\mathcal{F}_C$ is modeled as
\begin{equation}\label{Fcomp}
    \mathcal{F}_C \approx 10\,\mathrm{\mu mol\,m^{-2}\,s^{-1}}\, \mathcal{G}(T_W),
\end{equation}
where we have introduced the auxiliary function
\begin{equation}\label{AuxF}
  \mathcal{G}(T_W) \equiv  \exp\left[13.6\left(1-\frac{T_0}{T_W}\right)\right],
\end{equation}
with $T_0 \approx 289$ K representing the global surface temperature of Earth's oceans.\footnote{\url{https://www.ncdc.noaa.gov/sotc/global/201913}} The constant of proportionality in (\ref{Fcomp}) has been chosen as it represents the compensation flux for phytoplankton in Earth's oceans within a factor of $\sim 2$ \citep{NS91,Mar04,ML06,RD10}. By solving for $\mathcal{F}(z) = \mathcal{F}_C$, we are now equipped to calculate the compensation depth $\mathcal{Z}_\mathrm{CO}$ as a function of both the stellar temperature and ocean temperature.

In Fig. \ref{FigCDepG}, the photon flux normalized by the compensation flux is plotted as a function of the depth $z$ for the idealized case delineated in Sec. \ref{SSecPrelim}, where the star is at the substellar point and the attenuation in water is assumed to be minimal. The left panel corresponds to Planet G, while the right panel depicts the results for Planet M. By inspecting both panels, we find that $\mathcal{Z}_\mathrm{CO}$ decreases with the temperature along expected lines. The physical reason for this trend is that the increase in the rate of respiration outpaces that of photosynthesis when the temperature is elevated, thereby ensuring that the location at which the two processes balance each other is shifted closer to the surface of the ocean, i.e., leading to a reduction in $\mathcal{Z}_\mathrm{CO}$.

\begin{figure*}
$$
\begin{array}{cc}
  \includegraphics[width=7.1cm]{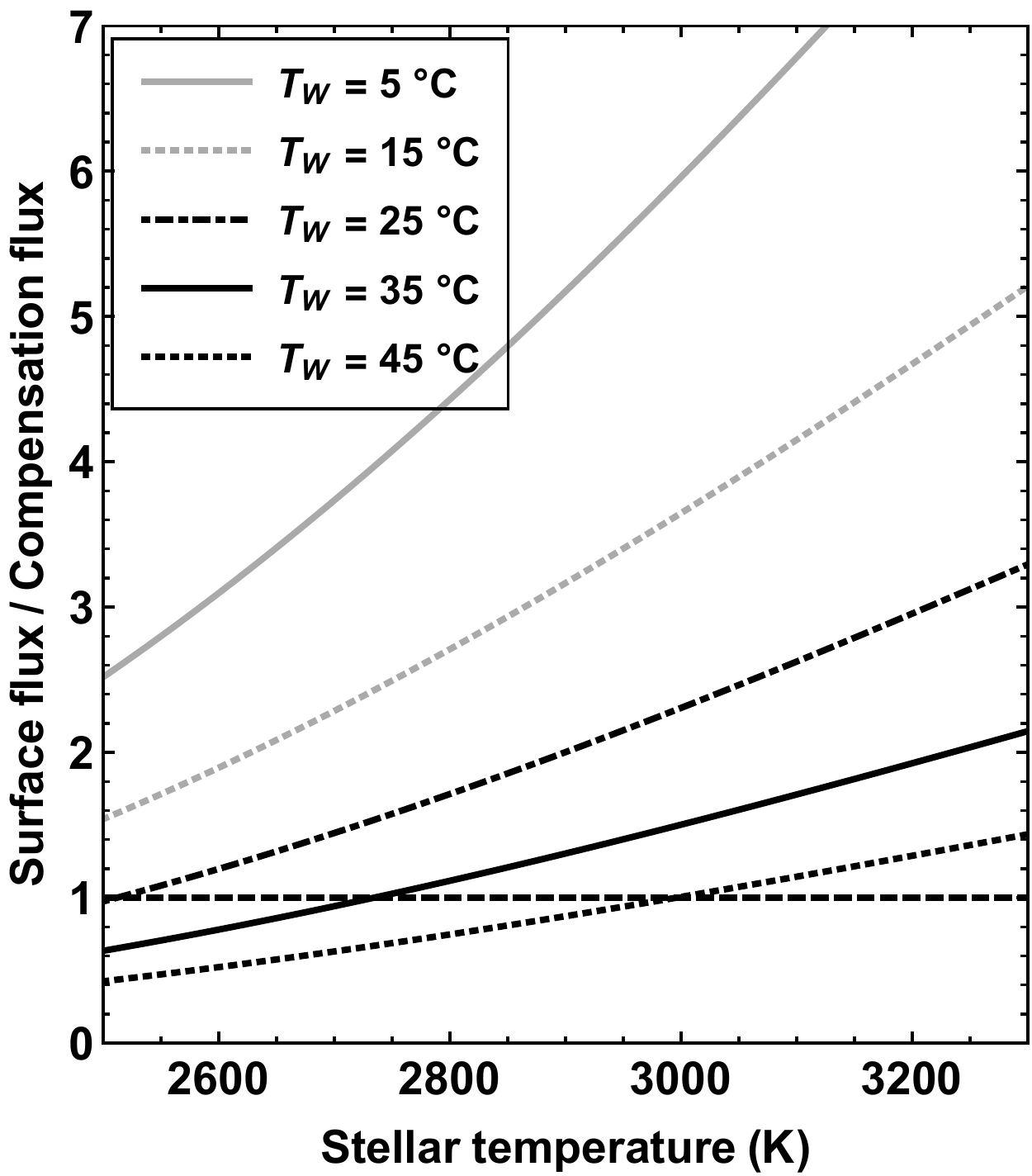} & 
  \includegraphics[width=7.5cm]{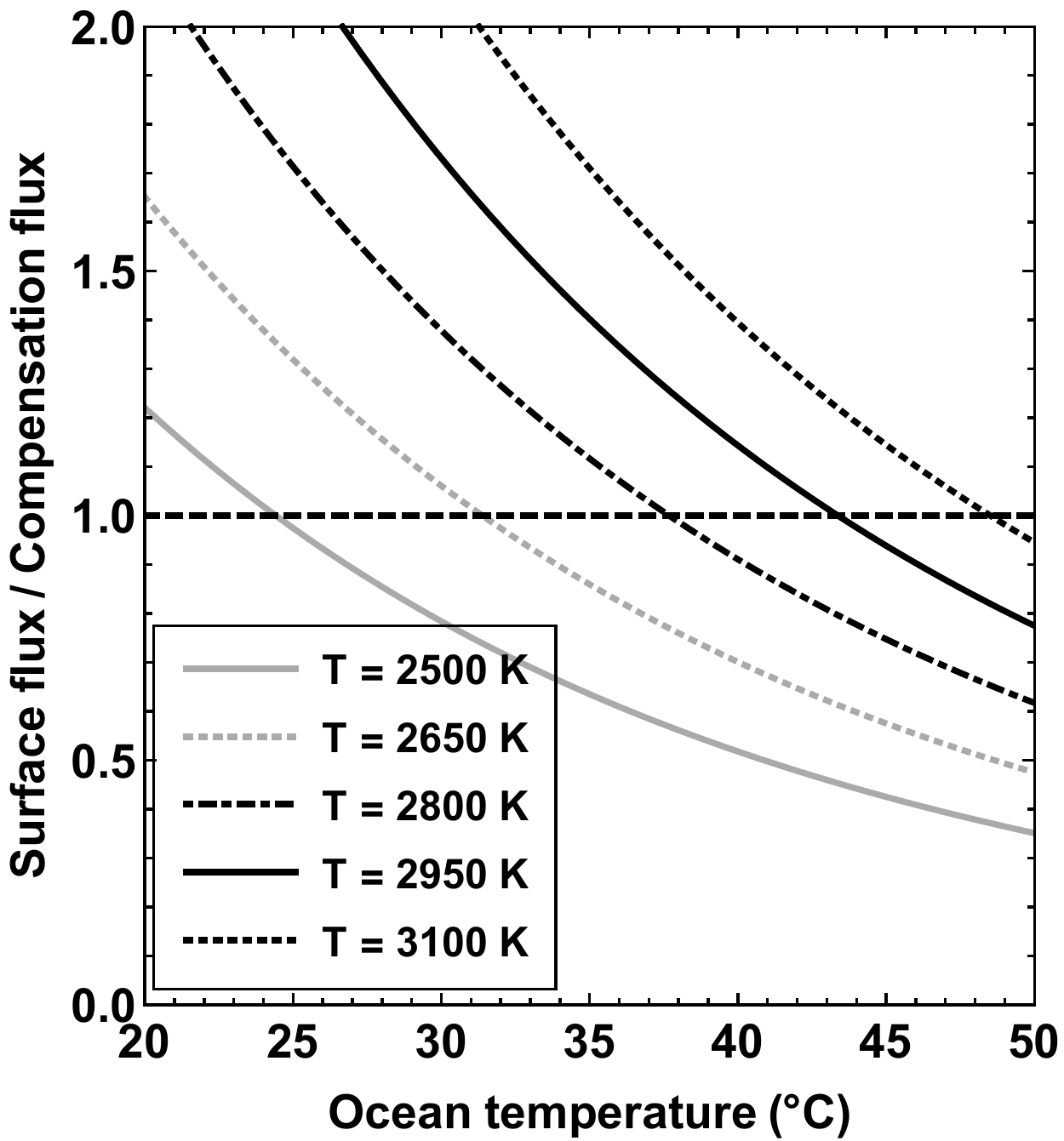}\\
\end{array}
$$
\caption{In both panels, the ratio of the photon flux at the surface to that of the compensation flux (denoted by $\zeta$) is depicted. Regions lying below the horizontal dashed line are relatively unlikely to host Earth-like biota in the oceans. Left panel: variation of $\zeta$ with stellar temperature (in K) for different ocean temperatures. Right panel: variation of $\zeta$ with ocean temperature (in $^\circ$C) for different stellar temperatures.}
\label{FigSurf}
\end{figure*}

We observe that the ocean temperature exerts a fairly significant effect on the magnitude of $\mathcal{Z}_\mathrm{CO}$ for both worlds. As far as Planet G (orbiting a solar twin) is concerned, the compensation depth changes from  $\mathcal{Z}_\mathrm{CO}^{(I)} \approx 300$ m at $T_W = 5$ $^\circ$C to $\mathcal{Z}_\mathrm{CO}^{(I)} \approx 130$ m at $T_W = 45$ $^\circ$C. On the other hand, when we consider Planet M, situated around a late-type M-dwarf closely resembling TRAPPIST-1, the compensation depth morphs from $\mathcal{Z}_\mathrm{CO}^{(I)} \approx 26.5$ m at $T_W = 5$ $^\circ$C to $\mathcal{Z}_\mathrm{CO}^{(I)} \approx 3.5$ m at $T_W = 45$ $^\circ$C. Hence, for the idealized case studied in this figure, we predict that the ocean temperature might cause $\mathcal{Z}_\mathrm{CO}$ to change by nearly an order of magnitude for Planet M; the variation associated with Planet G is smaller, but still non-negligible.

Fig. \ref{FigCDepM} is analogous to that of Fig. \ref{FigCDepG}, except that we consider the so-called global average case described at the end of Sec. \ref{SSecPrelim} in lieu of the idealized scenario. When it comes to Planet G (left panel), the compensation depth evolves from $\mathcal{Z}_\mathrm{CO}^{(A)} \approx 24$ m at $T_W = 5$ $^\circ$C to $\mathcal{Z}_\mathrm{CO}^{(A)} \approx 8.5$ m at $T_W = 45$ $^\circ$C. However, a striking result is manifested vis-\`a-vis Planet M (right panel). At $T_W = 5$ $^\circ$C, we obtain $\mathcal{Z}_\mathrm{CO}^{(A)} \approx 3$ m, but we end up with $\mathcal{Z}_\mathrm{CO}^{(A)} = 0$ at $T_W = 45$ $^\circ$C. The null value arises because the temperature elevates the compensation point to such an extent that it overshoots the incident photon flux at the ocean's surface.\footnote{We reiterate that our analysis deals with Earth-like biota, and the results are not necessarily applicable to putative oxygenic photoautotrophs in the oceans of M-dwarf exoplanets.} In fact, we determine that $\mathcal{F}_0 \equiv \mathcal{F}^{(A)}(z=0) < \mathcal{F}_C$ is fulfilled when $T_W > 24$ $^\circ$C, implying that ocean temperatures above this value appear to be relatively unsuitable for supporting phytoplankton-like biota on tidally locked Earth-analogs orbiting stars akin to TRAPPIST-1. 

Motivated by the above finding, we define $\zeta \equiv \mathcal{F}_0/\mathcal{F}_C$ and study the regimes in which $\zeta < 1$ is valid. This criterion enables us to gauge the conditions under which Earth-like oxygenic photoautotrophs may have a low likelihood of existing. We only tackle the global average case herein, as it permits $\zeta < 1$ to occur in the parameter space. From examining Fig. \ref{FigSurf}, where the results are depicted, it is apparent that some tidally locked late-type M-dwarf exoplanets might be incapable of hosting phytoplankton-like biota. In particular, for the upper bound of $T_W = 50$ $^\circ$C, we surmise that stars with $T < 3150$ K can be ruled out in this category. Thus, if the oceans are sufficiently warm, tidally locked Earth-analogs around late-type M-dwarfs could encounter difficulties in sustaining marine photosynthetic organisms analogous to modern Earth. 

Lastly, before proceeding further, there is one other point worth mentioning. As the depth of the photosynthesis zone grows more shallower, whether it be due to oceanic temperature or stellar spectral type, the photoautotrophs are expected to live closer to the surface. In doing so, they incur a greater risk of damage by ultraviolet radiation and energetic particles from flares and superflares, the latter of which could deposit high doses \citep{LL17,YMA,At20}. However, experiments and numerical models suggest that hazes (in)organic films \citep{CM98,EV18,LL19}, along with biogenic ultraviolet screening compounds and evolutionary adaptations \citep{CK99,ALO}, may suffice to protect them.

\subsection{Critical depth}

In order to calculate the critical depth ($\mathcal{Z}_\mathrm{CR}$), a number of different formulae have been delineated in the literature \citep{Sver53,Kirk11,Mid19}. Most of the simpler models reduce to \citep[equation 9.7]{FR07}:
\begin{equation}
    K \mathcal{Z}_\mathrm{CR} \approx \frac{\Gamma_P}{\Gamma_R},
\end{equation}
but they are correct only in the limiting case of $K = \mathrm{const}$, which is manifestly invalid. The generalization of the above equation was adumbrated in \citet{LL20}, who eventually obtained
\begin{equation}\label{CritD}
  \mathcal{Z}_\mathrm{CR} \approx \left(\frac{\Gamma_R}{\Gamma_P}\right)^{-1} \frac{\int_{\lambda_\mathrm{min}}^{\lambda_\mathrm{max}} \left[\mathcal{N}_0(\lambda)/K(\lambda)\right]\,d\lambda}{\int_{\lambda_\mathrm{min}}^{\lambda_\mathrm{max}} \mathcal{N}_0(\lambda)\,d\lambda}.
\end{equation}
It is, however, necessary to recognize that $\Gamma_R/\Gamma_P$ has an intrinsic temperature dependence, as seen from (\ref{GamRat}). Hence, we combine (\ref{CritD}) with (\ref{GamRat}), thereby yielding 
\begin{equation}\label{CritF}
  \mathcal{Z}_\mathrm{CR} \approx \frac{3.36 \times 10^{-2}}{\mathcal{G}(T_W)} \frac{\int_{\lambda_\mathrm{min}}^{\lambda_\mathrm{max}} \left[\mathcal{N}_0(\lambda)/K(\lambda)\right]\,d\lambda}{\int_{\lambda_\mathrm{min}}^{\lambda_\mathrm{max}} \mathcal{N}_0(\lambda)\,d\lambda},
\end{equation}
where the normalization has been adopted based on the global value for phytoplankton in Earth's oceans \citep[Chapter 4.3]{SG06}. The parameters pertaining to the ``A'' scenario are adopted for the sake of comparison with prior empirical studies. 

\begin{figure*}
$$
\begin{array}{cc}
  \includegraphics[width=7.3cm]{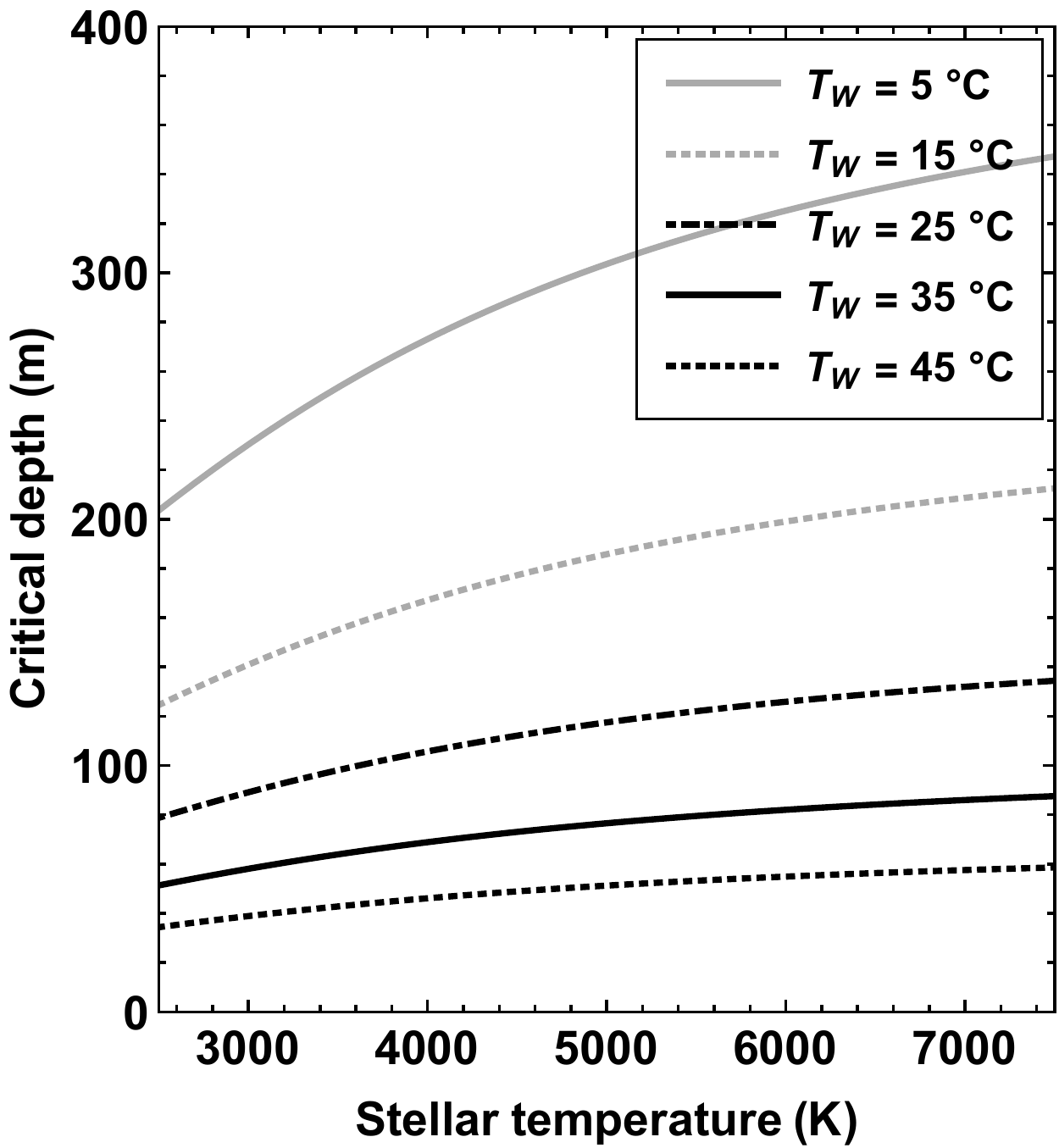} & 
  \includegraphics[width=7.5cm]{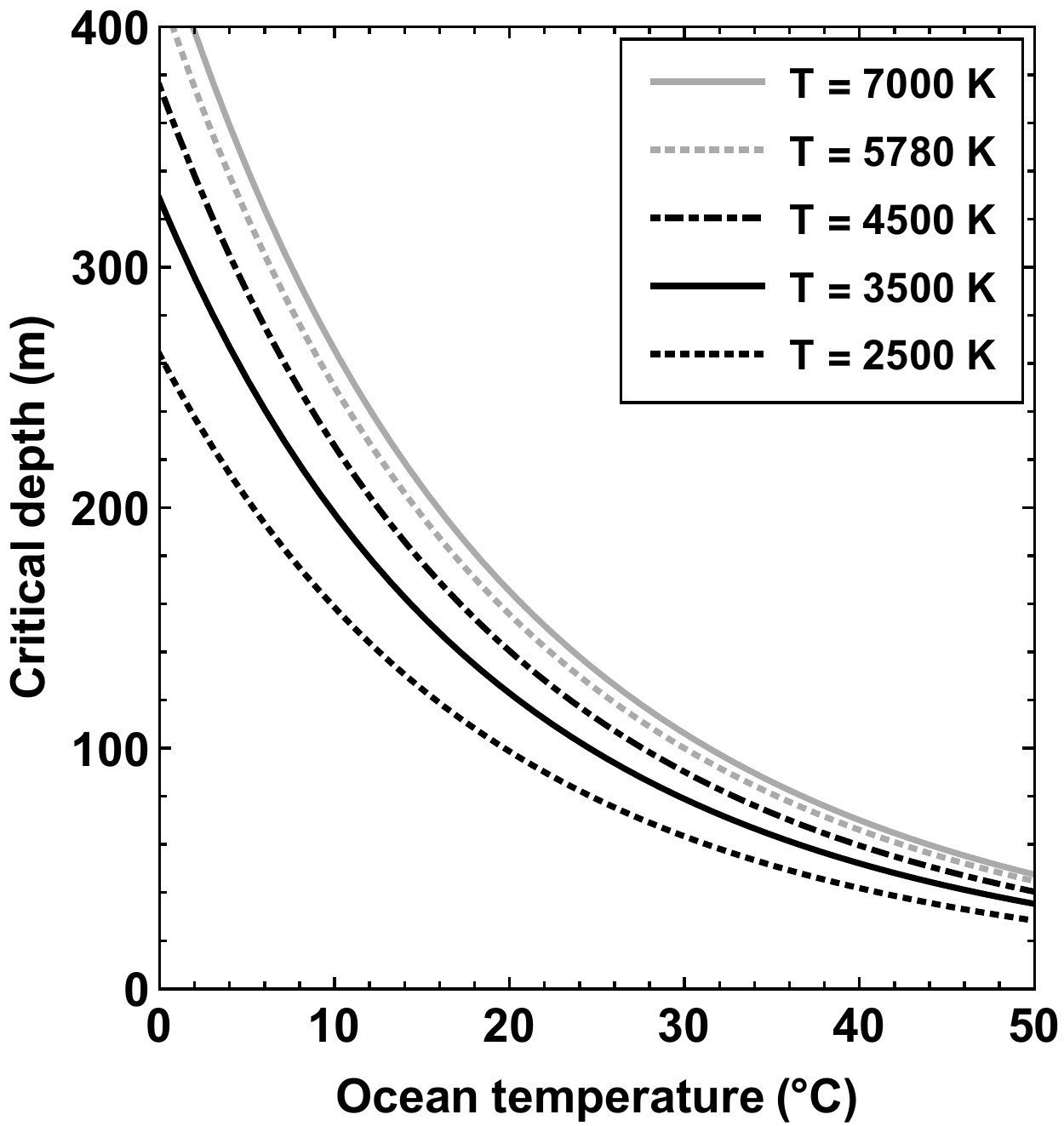}\\
\end{array}
$$
\caption{In both panels, the critical depth ($\mathcal{Z}_\mathrm{CR}$)---to wit, the location where the vertically integrated net growth rate is zero---is plotted. Left panel: variation of $\mathcal{Z}_\mathrm{CR}$ with stellar temperature (in K) depicted for different ocean temperatures. Right panel: variation of $\mathcal{Z}_\mathrm{CR}$ with ocean temperature (in $^\circ$C) illustrated for different stellar temperatures.}
\label{FigCritD}
\end{figure*}

The temperature dependence of the critical depth is illustrated in Fig. \ref{FigCritD}. Two points that emerge from scrutinizing this figure. From the left panel, we notice that the dependence on the stellar temperature is weak at any given ocean temperature. This result is consistent with \citet{LL20}, and is mostly attributable to the fact that net growth primarily occurs in the upper layers and thus compensates for the regions with $z > \mathcal{Z}_\mathrm{CO}$. However, when it comes to the ocean temperature, a much stronger variation of $\mathcal{Z}_\mathrm{CR}$ is discerned. As we cover the entire ocean temperature range considered herein, we find that $\mathcal{Z}_\mathrm{CR}$ changes by nearly an order of magnitude for any given stellar temperature (right panel). For instance, after we specify $T = T_\odot$, the critical depth evolves from $\mathcal{Z}_\mathrm{CR}^{(A)} \approx 416$ m at $T_W \approx 0$ $^\circ$C to $\mathcal{Z}_\mathrm{CR}^{(A)} \approx 45$ m at $T_W \approx 50$ $^\circ$C.

Therefore, it is conceivable that the ocean temperature plays a major role in regulating the critical depth on other worlds. In turn, this development suggests that $T_W$ also acts as a key determinant of phenomena analogous to phytoplankton blooms, which may constitute viable temporal biosignatures as noted earlier.

\subsection{Net primary productivity}\label{SSecNPP}

\begin{figure}
\includegraphics[width=7.5cm]{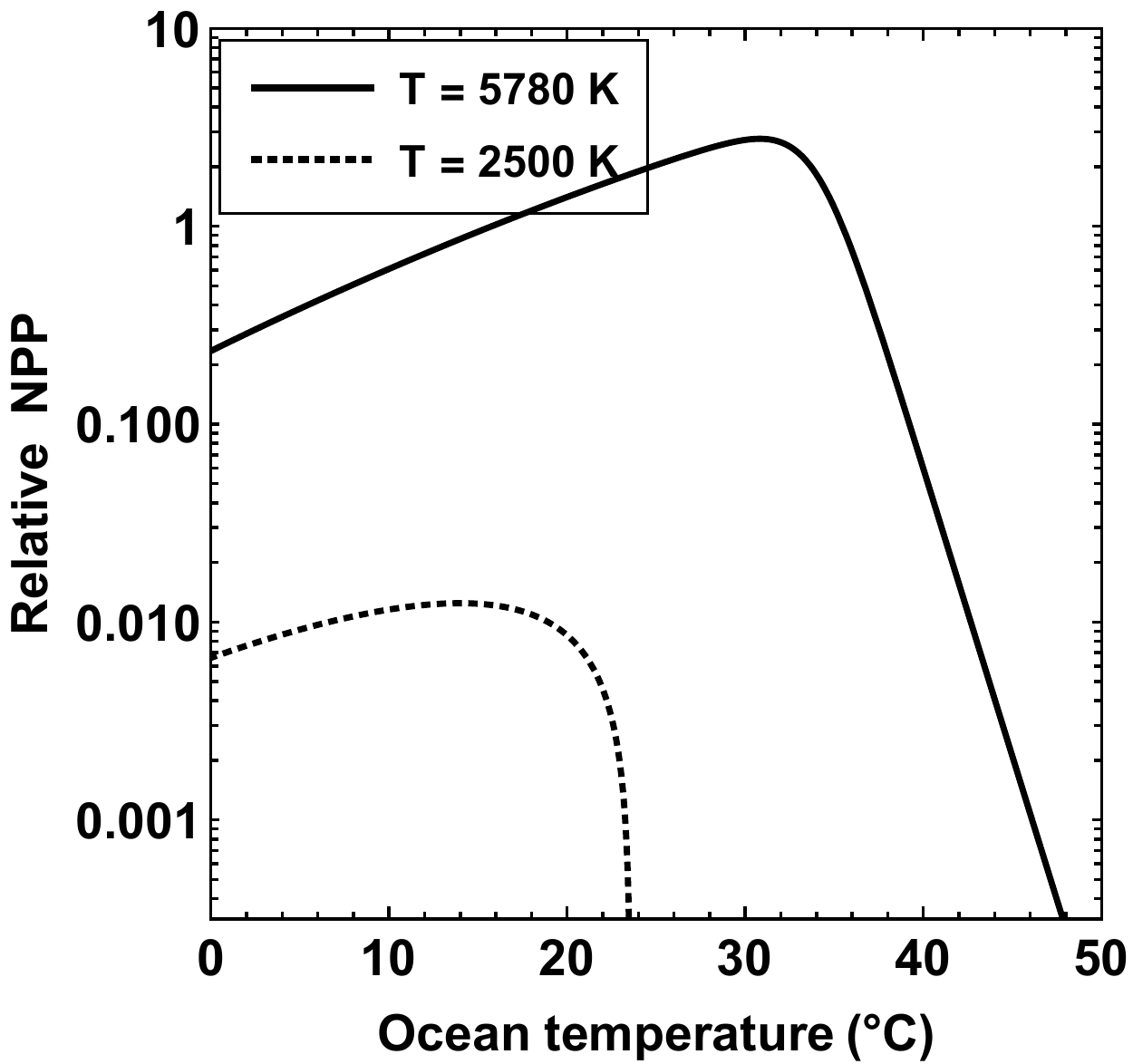} \\
\caption{The oceanic NPP relative to that of modern Earth as a function of the ocean temperature for Planet G (stellar temperature of $T = 5780$ K) and Planet M (stellar temperature of $T = 2500$ K). }
\label{FigNPP}
\end{figure}

\begin{table*}
\caption{Net primary productivity for the Earth-analogs as a function of the mean oceanic temperature}
\label{TabNPP}
\begin{tabular}{|c|c|c|}
\hline 
Ocean temperature ($^\circ$ C) & Relative NPP of Planet G & Relative NPP of Planet M \tabularnewline
\hline 
\hline 
$5$ & $0.4$ & $9 \times 10^{-3}$\tabularnewline
\hline 
$10$ & $0.6$ & $10^{-2}$\tabularnewline
\hline 
$15$ & $0.9$ & $10^{-2}$\tabularnewline
\hline 
$20$ & $1.4$ & $8 \times 10^{-3}$\tabularnewline
\hline 
$25$ & $2.0$ & $0$\tabularnewline
\hline 
$30$ & $2.7$ & $0$\tabularnewline
\hline
$35$ & $1.2$ & $0$\tabularnewline
\hline
$40$ & $6 \times 10^{-2}$ & $0$\tabularnewline
\hline 
$45$ & $2 \times 10^{-3}$ & $0$\tabularnewline
\hline 
\end{tabular}
\medskip

{\bf Notes:} The NPP is expressed in terms of the temporally averaged value associated with modern Earth's oceans, namely, $1.5 \times 10^{-2}\,\mathrm{g\, C\, m^{-2}\, h^{-1}}$. The NPP for these two Earth-analogs is calculated by deploying (\ref{NPPRes}). Planet G orbits a solar twin whereas Planet M is situated near a late-type M-dwarf akin to TRAPPIST-1; the other properties of the two planets are otherwise identical. 
\end{table*}

The NPP is arguably one of the most crucial and informative property of a biosphere as it quantifies the net amount of organic carbon synthesized via biological pathways after accounting for losses dues to respiration and other factors; we will express our results in units of g C m$^{-2}$ h$^{-1}$ for the NPP. The NPP is a reliable measure of the amount of organic C generated via photosynthesis, as the latter constitutes the dominant carbon fixation pathway \citep{Berg11,Knoll,Jud17,BPM18}. A wide spectrum of models have been developed to model NPP, and comprehensive reviews can be found in \citet{BF97} and \citet[Chapter 9]{FR07}. 

We make use of \citet[Equation 3]{FB98} to calculate the NPP, because this outwardly simple expression accounts for a number of environmental factors:
\begin{equation}\label{NPPDef}
    \mathrm{NPP} = C_\mathrm{sur} \cdot \mathcal{Z}_\mathrm{CO}^{(A)} \cdot f(\mathrm{PAR}) \cdot P_\mathrm{opt}(T_W),
\end{equation}
where $C_\mathrm{sur}$ is the chlorophyll concentration at the surface, $f(\mathrm{PAR})$ embodies the fraction of the water column up to $\mathcal{Z}_\mathrm{CO}^{(A)}$ where photosynthesis is light saturated, and $P_\mathrm{opt}(T_W)$ is the optimal carbon fixation rate. There exists, however, an inherent crucial subtlety that needs to be spelt out here. In canonical versions of the above formula, $\mathcal{Z}_\mathrm{CO}^{(A)}$ is replaced by the euphotic zone depth. However, as noted in \citet[pg. 237]{FB98}, the proper variable that ought to be deployed is the depth of the zone where positive NPP is feasible, which is congruent with the definition of the compensation depth. On Earth, the euphotic zone depth \citep{LWK07} and the compensation depth \citep{SJF42,Mid19} are roughly equal to one another, but the same relationship is not necessarily valid \emph{a priori} for other worlds; even on Earth, the reliability of the euphotic zone as a measure of the photosynthesis zone has been called into question \citep{Ban04,MLV14}.

The NPP will depend not only on the stellar and ocean temperatures but also on inherent biological factors such as $C_\mathrm{sur}$ that are spatially and temporally very heterogeneous. As the goal of the paper is to construct heuristic global estimates, we rewrite (\ref{NPPDef}) so that it yields the average global value for the Earth at $T = T_\odot$ and $T_W = T_0$ (i.e., the parameters for Earth). By adopting the normalization from \citet{FB98},\footnote{We note that some subsequent estimates for the oceanic NPP have revised the classic analysis of \citet{FB98} by $\mathcal{O}(10\%)$ \citep{WBS08}, but this has a minimal impact on both our subsequent qualitative and quantitative results.} we obtain
\begin{eqnarray}\label{NPPRes}
 && \mathrm{NPP} \approx 1.5 \times 10^{-2}\,\mathrm{g\, C\, m^{-2}\, h^{-1}} \left(\frac{\mathcal{Z}_\mathrm{CO}^{(A)}}{\mathcal{Z}_0}\right)  \nonumber \\
 && \hspace{0.5in} \times \,\mathcal{P}(T_W) \left(\frac{\mathcal{D}}{0.5}\right) \left(\frac{G(T)}{G(T_\odot)}\right),
\end{eqnarray}
where $\mathcal{Z}_0 \approx 19$ m represents the compensation depth calculated at the fiducial ocean temperature of $T_0$ using the methodology in Sec. \ref{SSecCoD}, while $\mathcal{D}$ denotes the fraction of time that a given location receives stellar illumination. For planets like Earth, we expect $\mathcal{D} \approx 0.5$ (i.e., equipartition of day and night) whereas tidally locked planets ought to have $\mathcal{D} \approx 1$ on the day side because they receive stellar radiation \emph{in perpetuo}. The auxiliary functions $G(T)$ and $\mathcal{P}(T_W)$ are defined as follows:
\begin{equation}\label{Gdef}
 G(T) =  \frac{\mathcal{F}_0}{\mathcal{F}_0 + \mathcal{F}_S}, 
\end{equation}
where $\mathcal{F}_S \approx 1.1 \times 10^3$ $\mu$mol m$^{-2}$ s$^{-1}$, and the stellar temperature is implicitly present via $\mathcal{F}_0$. 
\begin{eqnarray}
&& \mathcal{P}(T_W) = \left[ \frac{1 + \exp\left[\frac{E_h}{k_B}\left(\frac{1}{T_h} - \frac{1}{T_0}\right)\right]}{1 + \exp\left[\frac{E_h}{k_B}\left(\frac{1}{T_h} - \frac{1}{T_W}\right)\right]}\right] \nonumber \\
&& \hspace{0.6in} \times \exp\left[\frac{E_a}{k_B}\left(\frac{1}{T_0} - \frac{1}{T_W}\right)\right],
\end{eqnarray}
where $E_a \approx 0.74$ eV, $E_h \approx 6.10$ eV and $T_h \approx 34$ $^\circ$C are adopted for our putative biota from \citet[pg. 726]{BJ20}.\footnote{It goes without saying that all of these parameters exhibit substantive variation across species. For instance, the thermal performance curves for certain species of marine phytoplankton reveal optimal temperatures of $\sim 20$-$25$ $^\circ$C \citep{BRAF}, in which case $T_h$ is lowered by several $^\circ$C.} Here, we have constructed (\ref{NPPRes}) and (\ref{Gdef}) based on \citet[Section 2.4]{BBS05} and \citet[Equation 10]{BF97}, but one point of divergence is that a modified Sharpe–Schoolfield equation \citep{SD77,SSM81} was utilized as a proxy for $P_\mathrm{opt}(T_W)$, following \citet{BYD19,BJ20} in lieu of \citet[Equation 11]{BF97}, as the latter becomes invalid for $T_W > 30$ $^\circ$C. The precise expression for $P_\mathrm{opt}(T_W)$ for phytoplankton is challenging to accurately pin down, owing to the panoply of expressions used to model phytoplankton growth \citep{GMS17}. In consequence, a diverse array of functions, some exhibiting exactly opposite trends with temperature, have been employed for this purpose \citep[Figure 4]{MBPF}; see also \citet{GMS17}. Hence, the ensuing results should be interpreted with due caution.

We have presented the NPP for the two Earth-analogs (Planet G and Planet M) in Table \ref{TabNPP} and Figure \ref{FigNPP}, which were calculated by using the global average case as seen from (\ref{NPPDef}). There are several interesting results that emerge from inspecting these two items. We begin by considering Planet G (orbiting a solar twin) to gauge the role of $T_W$. We notice that the NPP increases with ocean temperature until $\sim 30$ $^\circ$C, but the growth is relatively modest. It is primarily driven by the rise in the rate of carbon fixation, as encapsulated by $\mathcal{P}(T_W)$, with temperature in this regime. In some controlled experiments and modeling where the temperature was steadily elevated, the photosynthetic capacity has been found to increase up to a point \citep{LBH14,PYD16,SBB17}.\footnote{It is important to recognize, however, that the variation of NPP with temperature is subject to much uncertainty due to the large number of coupled variables and nonlinear feedback mechanisms \citep{TO11,LVG15}.} As per our simple model, once the peak temperature of the thermal performance curve is exceeded ($T_\mathrm{pk}$), the rate of carbon fixation falls sharply thereafter, and consequently drives the steep decline in NPP when $T_W > 35$ $^\circ$C. 

Now, we turn our attention to Planet M around a late-type M-dwarf similar to TRAPPIST-1. For any fixed temperature, say $T_W = 5$ $^\circ$C, we notice that the NPP is lower than Planet G by roughly two orders of magnitude. The reasons for the diminished NPP are twofold: (i) the compensation depth is greatly reduced as pointed out in Sec. \ref{SSecCoD}, and (ii) the flux of PAR is corresponding lower at the surface, thereby making the last term on the right-hand-side of (\ref{NPPRes}) smaller than unity. The next major feature we notice is that the NPP vanishes at $T_W \sim 24$ $^\circ$C. This result is a direct consequence of the fact that the compensation depth becomes zero above a threshold temperature for reasons explained in Sec. \ref{SSecCoD}. Hence, tidally locked exoplanets around late-type M-dwarfs may evince a low likelihood of large-scale carbon fixation by phytoplankton-like biota. Needless to say, the NPP is not anticipated to be zero \emph{sensu stricto}, because anoxygenic photoautotrophs are capable of carbon fixation by definition \citep{Ko07,SB13}, and so are many microbial taxa in the deep biosphere \citep{OSK11,EBC12,MP14,CPS17}.

\section{Limitations of the model}\label{SecModLim}
It is worth emphasizing at the outset that the productivity of biospheres is constrained by a number of factors including water, electron donors, temperature, PAR flux and nutrients \citep{LL21}. Our analysis tackles the modulation of the productivity of biospheres by PAR and ambient ocean temperature \emph{ceteris paribus}. In doing so, we follow the likes of \citet{LCPH,LM19} in setting aside the constraints imposed by the access to nutrients and some of the other variables.

In the case of Earth's terrestrial (land-based) NPP, the NPP for $> 80\%$ of the area is limited by water and temperature \citep{CR98}. In contrast, Earth's oceanic NPP---both the globally averaged value and the spatiotemporal variations---is governed by the prevalence of nutrients, especially the ultimate limiting nutrient phosphate \citep{Ty99,Fil08,SB13}. Ocean planets, in particular, may be impacted due to their potentially lower rates of weathering and delivery of nutrients to the oceans \citep{WP13,LM18,KF18,LiMa}. The key caveat in this paper, therefore, is that the oceanic NPP is not constrained by nutrients, but is instead regulated the two factors adumbrated in the preceding paragraph. The ensuing results might comprise upper bounds for the NPP because the abundance and distribution of nutrients could introduce additional limits.

As we shall demonstrate in Sec. \ref{SSecNPP}, the oceanic NPP for tidally locked Earth-like exoplanets around late-type M-dwarfs is severely constrained by the paucity of PAR photons, and is orders of magnitude smaller than Earth's oceanic NPP. Hence, the prior assumption might not pose a major problem for these worlds because the most dominant bottleneck on the oceanic NPP may prove to be the PAR flux. However, when it comes to Earth-analogs around Sun-like stars, PAR flux is not a major limiting factor and the thermal effects on NPP might become prominent only at high temperatures. Thus, a brief discussion of nutrient limitation and how it could impact the oceanic NPP of other worlds is apropos.

The first and foremost point that needs to be appreciated is that modeling nutrient limitation even on Earth is a complicated endeavor. The reason is that the nutrient concentration in the ocean depends on a variety of factors such as the remineralization efficiency \citep{KS17,LS18}, hydrothermal activity \citep{WFM96}, submarine weathering \citep{Hao20}, and mineral solubility in seawater \citep{Derr15}, to name just a few. Moreover, each of these quantities has fluctuated over time and witnessed shifts in magnitude, sometimes up to a factor of $\sim 10$ as may have occurred vis-\`a-vis the remineralization efficiency during the Ediacaran period \citep{LSJ20}. For these reasons, theoretical models for the biogeochemical cycles of the bioessential elements have yielded very different results \citep{Len20}; see also \citet{Hao20} for an exposition of this issue.

As the prior discussion suggests, there are numerous mechanisms that control the nutrient concentration in oceans. In consequence, it is not inconceivable that some Earth-like planets could bypass or mitigate the issue of nutrient limitation. Geological processes that have been proposed hitherto for counteracting the nutrient deficiency to varying degrees include elevated nutrient upwelling \citep{LL18,OJA,SOK20}, submarine basalt weathering \citep{SRI20} and serpentinization \citep{POL20}. As noted earlier, we will presume hereafter that nutrient abundance is not the chief limitation, perhaps via some of the above channels coming into play. To reiterate, we suppose that either photon flux and/or temperature act to throttle the productivity. We will demonstrate hereafter that these factors become exceedingly important for planets around late-type M-dwarfs and/or with high ambient ocean temperatures; in particular, the NPP might become orders of magnitude smaller relative to Earth.

In relation to the preceding points, we note that the constraints imposed by the ambient photon flux, temperature and nutrients do not act independently of one another. In fact, a multitude of experiments and field studies have established that these environmental parameters are non-linearly coupled to one another \citep{ETK16,SMS16,GMS17,TAK17,MLC18}. For instance, the value of $E_a$ introduced previously may vary significantly in some species depending on the availability of nitrogen \citep{MLC18} and the ocean temperature \citep{MBM20}. While such effects are indubitably important, they are not well understood even on Earth and exhibit considerable intra- and inter-species variability. Hence, given that the implicit goal of this paper was to construct heuristic models that provide rough estimates for future observations and modeling, we have not taken these subtle processes into account.

Lastly, in our subsequent analysis, we will draw upon the basic physiological properties of the dominant phytoplankton species on Earth. While this line of reasoning is undoubtedly parochial, we note that Earth-based organisms are commonly used as proxies in numerous astrobiological contexts \citep{MCK17}, ranging from extremophiles and microbial ecosystems in the oceans of icy moons \citep{CH01,RM01,CKB17,MCK17,MAB19,LMa19} to the limits of complex multicellular life on exoplanets \citep{SVS17,SRO19,Li20,Ram20}. Furthermore, the choice of phytoplankton as putative biota is motivated by the fact that they are the major source of carbon fixation in the oceans of modern Earth \citep{Har86,Fal04,CKT05,Rav09,UCG10}. Hence, by utilizing the prior framework, we are now equipped to analyze the prospects for Earth-like aquatic photosynthesis on other worlds characterized by different ocean temperatures.

\section{Discussion}\label{SecDisc}
We will discuss some of the implications of our work in connection with mapping the trajectories of the Earth as well as tidally locked M-dwarf exoplanets.

\subsection{Potential future evolution of Earth}
We begin by tackling the ramifications of the preceding analysis for the Earth's aquatic biosphere, with respect to its potential future. 

Before doing so, it is worth briefly highlighting the inherent spatiotemporal variability of Earth's oceanic NPP. To begin with, let us recall that a global sea surface temperature (SST) of $T_0 \approx 16$ $^\circ$C was chosen herein based on satellite data. However, in reality, the SST of Earth is characterized by distinct heterogeneity, ranging from $35$ $^\circ$C to below-freezing temperatures.\footnote{\url{https://earthobservatory.nasa.gov/global-maps/MYD28M}} Moreover, the Earth's NPP is modulated by the access to not only light and temperature (both of which are present in our model) but also nutrients \citep{BBS05}; the latter may play a crucial role as noted in Sec. \ref{SecModLim}. Collectively, these factors engender variations in the oceanic NPP across both the spatial and temporal domains \citep{WBS08}, sometimes by roughly an order of magnitude. Thus, we reiterate that our model only seeks to extract globally averaged values for the relevant variables from a heuristic standpoint.

There is a sharp downswing in NPP shortly after the peak temperature $T_\mathrm{pk}$ is attained, which becomes evident upon inspecting Fig. \ref{FigNPP}. While there are grounds for contending that $T_\mathrm{pk} \sim 30$ $^\circ$C \citep{BJ20}, this matter is admittedly not conclusively settled. Now, let us suppose that the Earth's temperature was raised by $\sim 10$ $^\circ$C abruptly. In large swathes of the ocean, it is conceivable that $T_W > T_\mathrm{pk}$, thereby triggering a sharp downswing in the NPP in these regions. In turn, given that phytoplankton are the foundation of oceanic food webs and trophic interactions \citep{BH99,Val15,Kir18}, this rapid decline in NPP ought to have adverse consequences for marine ecosystems and could thus potentially drive large-scale extinctions of marine biota.

As the Sun continues to grow brighter, the surface temperature will also increase commensurately because of the greenhouse effect until the Earth is eventually rendered uninhabitable \citep{CK92,GW12,RCOW}. Based on \citet[Section 3.1]{WT15}, we note that a global temperature of $312$ K is predicted when the solar luminosity is $1.1$ times the present-day value. By utilizing \citet[Equation 1]{Go81}, the stellar luminosity associated with this temperature is expected to occur $\sim 1.2$ Gyr in the future. It is important to note, however, that climate models do not fully agree on the critical flux at which the greenhouse state is likely to be activated, implying that a timescale of $< 1$ Gyr ought not be ruled out \citep{GRZ13,LFC13,KCK15,PSM16,WSK17}.  

If we suppose that the global ocean temperature tracks the average surface temperature, the above analysis suggests that $T_W \sim 39$ $^\circ$C would occur $\sim 1.2$ Gyr hereafter. After examining Fig. \ref{FigNPP}, we find that the oceanic NPP at this $T_W$ might be $< 10\%$ of modern Earth. Due to the diminished NPP, eventual depletion of atmospheric O$_2$ is plausible for reasons adumbrated in Sec. \ref{SSecMDExo}, namely, when the sinks for oxygen outpace the sources. A decline in atmospheric O$_2$ could, in turn, drive the extinction of motile macroscopic organisms, as their long-term survival customarily necessitates oxygen levels $\sim 10\%$ of their present value \citep{CGZM,Wil09,ZC16,RPO16}.\footnote{In contrast, relatively sessile animals, such as the demosponge \emph{Halichondria panicea} \citep{MWJ14}, are capable of surviving at oxygen levels around $2$-$3$ orders of magnitude smaller than today \citep{SKG15,LK18}.} Thus, \emph{in toto}, the biosphere is unlikely to exhibit the same complexity as that of present-day Earth: this qualitative result is broadly consistent with earlier predictions by \citet{OMJ13,OMJ14}.

\subsection{Tidally locked M-dwarf exoplanets}\label{SSecMDExo}
We turn our attention to Planet M, i.e., the putative tidally locked exoplanet around a late-type M-dwarf similar to TRAPPIST-1.

It is instructive to compare our results against prior analyses of related topics. \citet[Table A9]{WoRa} calculated the oceanic NPP, albeit at a fixed depth of $10$ m using a simple model based on the photon flux, and estimated that it was $\sim 5$ times lower for an Earth-analog around an M0 star. In a similar vein, \citet{LCPH} and \citet{LM19} employed simple models for the NPP that were linearly proportional to the incident photon flux and determined that planets orbiting late-type M-dwarfs are unlikely to host biospheres with the same NPP as modern Earth and build up atmospheric O$_2$ to detectable levels. Thus, by and large, our work maintains consistency with earlier studies, but it has taken several other environmental and physiological variables into account that were missing in previous analyses.

We have previously calculated that the NPP for Planet M is, at most, only a few percent of the Earth's current oceanic NPP. Hence, because of the low NPP, unless the burial efficiency of organic carbon is unusually high, it seems likely that the flux of O$_2$ contributed by oxygenic photosynthesis will be correspondingly small. Hence, it ought to become more feasible for the sinks of atmospheric O$_2$ (e.g., continental weathering and volcanic outgassing) to dominate this source (which is a major player on Earth). The end result is that O$_2$ has a low likelihood of accumulating to detectable levels in the atmosphere \citep{CK17}. 

This potential effect has two consequences in turn. First, O$_2$ has been conjectured to be an essential prerequisite for complex life insofar as metabolism is concerned \citep{Kno85,McK96,CGZM,LL21}, at least up to a certain threshold after which oxygen toxicity may set in \citep{Li20}. Hence, the evolution of complex life, and potentially technological intelligence, might be suppressed on this category of worlds. Second, and more importantly, the absence of detectable atmospheric O$_2$ or O$_3$ for the aforementioned reasons despite the existence of a biosphere is an archetypal example of a ``false negative'' that can hinder or complicate the search for extraterrestrial life \citep{ROS17,MRA18}. 

\subsection{Build-up of atmospheric oxygen on ocean worlds}
It is worth quantifying the above qualitative treatment to gain further insights for ocean planets that are otherwise akin to present-day Earth. We will adopt the prescription laid out in \citet{LCPH}. We begin the analysis by noting that Earth's current oceanic NPP translates to an O$_2$ production flux of $5 \times 10^{-4}$ mol m$^{-2}$ h$^{-1}$, because the simplified reaction scheme for oxygenic photosynthesis takes the form
\begin{equation}\label{PhotOx}
    \mathrm{CO_2} + 2 \mathrm{H_2O} \, \xrightarrow[\text{pigments}]{h \nu}\, \mathrm{CH_2O} + \mathrm{H_2O} + \mathrm{O_2},
\end{equation}
where CH$_2$O embodies the synthesis of organic compounds, and H$_2$O appears in both sides of the equation as reactant and product, respectively. However, only a minuscule fraction of this O$_2$ is deposited in the atmosphere, since the vast majority is consumed by respiration and oxidative decay. If we denote the burial fraction by $\phi$, the flux of O$_2$ produced is then given by
\begin{equation}
\dot{S} \sim 1.5 \times 10^{-6}\,\mathrm{mol\,m^{-2}\,h^{-1}}\,\left(\frac{\mathrm{NPP}}{\mathrm{NPP}_\oplus}\right)\left(\frac{\phi}{\phi_\oplus}\right),
\end{equation}
where $\dot{S}$ is the O$_2$ flux generated from organic carbon burial, $\mathrm{NPP}_\oplus$ is the globally averaged oceanic NPP of the Earth, and $\phi_\oplus \approx 3 \times 10^{-3}$ is the fraction of organic carbon (fixed by photosynthesis) subjected to burial on present-day Earth \citep{Holl02,LCPH}. In order for O$_2$ to build up on anoxic worlds, the above source must exceed the primary sink, namely, reducing gases arising from a mixture of surface and submarine volcanism, metamorphism and serpentinization to name a few \citep{CK17}. We introduced $\dot{D}$, the depletion flux of O$_2$ associated with reducing gases, and specify a fiducial value of $\dot{D}_\oplus \sim 1.3 \times 10^{-6}\,\mathrm{mol\,m^{-2}\,h^{-1}}$ for modern Earth \citep{CK17}. The criterion for O$_2$ accumulation in the atmosphere is thus expressible as $\dot{S} > \dot{D}$, which simplifies to
\begin{equation}\label{NPPcond}
  \mathrm{NPP}  > 0.9 \mathrm{NPP}_\oplus \left(\frac{\phi}{\phi_\oplus}\right)^{-1} \left(\frac{\dot{D}}{\dot{D}_\oplus}\right).
\end{equation}
Hence, the above relation suggests that the oceanic NPP must be close to its present-day value in order for the build up of atmospheric O$_2$ to potentially take place, if all other parameters are held fixed. In contrast, if the burial of carbon is very efficient or the flux of reducing gases is extremely low, even a NPP that is much smaller than that of modern Earth may support the accumulation of atmospheric O$_2$. We run into an immediate difficulty here since both $\phi$ and $\dot{D}$ are not tightly constrained for Earth-like worlds in general.

However, if we interpret Earth-analogs to include only those worlds with all geochemical parameters similar to Earth, we can make headway. In such cases, (\ref{NPPcond}) reduces to the simpler $\mathrm{NPP}  > 0.9 \mathrm{NPP}_\oplus$. By comparing this criterion with Fig. \ref{FigNPP}, it is possible to deduce the conditions that permit the build up of atmospheric O$_2$ if only light and temperature constitute the sole limiting factors (cf. the next paragraph). For Earth-analogs around solar twins, we find that a temperature range of $\sim 15$-$35$ $^\circ$C might permit the build up of atmospheric O$_2$. At much higher and lower temperatures, the NPP is accordingly diminished, owing to which the reducing gases could overwhelm the O$_2$ generated from carbon burial. When we consider tidally locked Earth-analogs around late-type M-dwarfs, Fig. \ref{FigNPP} and Table \ref{TabNPP} suggest that the NPP would be only a few percent of Earth's current oceanic NPP. In that event, it might not be feasible for the accumulation of atmospheric O$_2$ to occur, as per the simplified formalism we have adopted.

\subsection{Observational tests for the future}
It is helpful to examine the prospects for testing our results by means of future observations at this juncture. A number of publications mentioned at the beginning of Sec. \ref{SSecMDExo} have already propounded strategies to gauge whether the stellar spectral type affects the NPP and the accompanying rise in atmospheric O$_2$ levels. The basic idea is to search for correlations between the spectral type of the host star on the one hand and the presence/absence of O$_2$ on the other. However, these putative correlations need to be weighed carefully because of the presence of major sources and sinks of O$_2$ not prevalent on Earth; for instance, the abiotic build-up of O$_2$ may be driven by electromagnetic radiation \citep{WP14,LB15,HSS15,KWF18} or its depletion may be effectuated by intense stellar winds and space weather events \citep{GGD17,DLMC,DHL17,DJL18,DLY18,DHL19,DJL20,ABC20}.

Therefore, we will restrict ourselves to assessing the observational implications insofar as worlds with varying ocean temperature are concerned. A scrutiny of Fig. \ref{FigNPP} reveals that a fairly steep decline in the NPP is potentially anticipated above a certain cutoff temperature. On the other hand, the surface density of chlorophyll ($C_\mathrm{sur}$) appearing in (\ref{NPPDef}) might not be affected to the same degree; in fact, we have held it fixed for the sake of simplicity. Hence, at least in principle, the detection of photoautotrophs ought to be feasible via the photosynthetic red edge \citep{STSF}, especially in the event that the organisms cover a large fraction of the surface \citep{OK19}. 

If we can therefore sample enough planets and discern a critical mean ocean temperature (the surface temperature might comprise a rough proxy for $T_W$) for a particular spectral type above which no biogenic O$_2$ and O$_3$ are detected but a tangible photosynthetic red edge is identified, such a distinct correlation could provide an avenue for falsifying our hypothesis. However, we caution that this strategy is not easily implementable in the near-future because it necessitates access to a sufficiently large sample of worlds with confirmed reliable biosignatures, oceans, and surface temperature measurements \citep{Kal17,SKP18,FAD18,Mad19}.

\section{Conclusions}\label{SecConc}
In this paper, we investigated how the ambient ocean temperature $T_W$ and the spectral type of the host star may influence the characteristics of aquatic biospheres on Earth-like worlds, albeit under a set of key assumptions that were expounded in Sec. \ref{SecModLim}.

In spite of the underlying simplifications and the ensuing limitations, there are several new results that were presented in this work, of which some of the major ones are outlined below and described in more detail later.
\begin{itemize}
    \item The compensation depth and critical depth is calculated as a function of the mean ocean temperature for Earth-analogs around a late-type M-dwarf (Planet M) and a Sun-like star (Planet G); it should be noted that Planet M is modeled as being tidally locked.
    \item The \emph{vertically integrated} average oceanic NPP is estimated for Planet G and Planet M. In other words, the procedure for determining the oceanic NPP as a function of the spectral type and the ocean temperature was explicated.
    \item The criterion for the accumulation of oxygen (O$_2$) in the atmospheres of Planets G and M is derived; this criterion is dependent not only on the spectral type but also on the ocean temperature.
\end{itemize}

We began by estimating the compensation depth and critical depth, as they serve to quantify the depths at which the net growth rate and vertically integrated net growth rate become zero, respectively. We showed that the ocean temperature has a relatively moderate influence on the compensation depth for an Earth-analog around a solar twin, as $\mathcal{Z}_\mathrm{CO}$ varies only by a factor of $< 3$. In contrast, when it comes to an Earth-like tidally locked world orbiting a late-type M-dwarf akin to TRAPPIST-1, we found that $T_W$ causes $\mathcal{Z}_\mathrm{CO}$ to vary by at least an order of magnitude. Furthermore, sufficiently warm oceans may preclude phytoplankton-like biota from existing on these worlds altogether. We calculated the critical depth and showed that it is sensitive to $T_W$, and varies by nearly an order of magnitude for the temperature range considered herein. 

Next, we examined the oceanic NPP of Planet G and Planet M as a function of the ocean temperature. The NPP constitutes one of the most vital metrics for a biosphere, and it has practical consequences that are delineated in the next paragraph. This calculation entailed the estimation of several variables, of which a few have not been robustly determined as empirical functions of $T_W$. Bearing this caveat in mind, we found that the NPP on Planet G was not very sensitive to $T_W$ until it exceeded a certain threshold after which the rate of carbon fixation dropped precipitously and drove a corresponding decline in the NPP. For the case of Planet M, the NPP was determined to be $\lesssim 1\%$ that of modern Earth, primarily on account of the shallowness of the photosynthesis zone in tandem with the lower PAR fluxes. When the ocean temperatures were raised sufficiently, the conditions for phytoplankton-like biota became untenable as noted in the earlier paragraph, and consequently resulted in the NPP approaching zero.

Lastly, we analyzed the ramifications of our work in the context of our planet as well as tidally locked Earth-like exoplanets orbiting late-type M-dwarfs. We discussed how an increase of $\sim 10$ $^\circ$C in the ocean temperature, such as what is expected to happen $\lesssim 1$ Gyr in Earth's future due to the growing solar luminosity, could radically transform the aquatic biosphere of Earth-analogs around G-type stars and diminish the NPP to $< 10\%$ of the Earth's current oceanic NPP in large swathes of the oceans. In a similar vein, we surmised that the aquatic biospheres of tidally locked Earth-like worlds around late-type M-dwarfs may evince NPPs that are $\lesssim 1\%$ of our planet's oceanic NPP today. If this prediction is correct, these worlds would be unlikely to accumulate atmospheric O$_2$---except in circumstances where they have much higher carbon burial and lower outgassing of reducing gases---but signatures of life are nonetheless potentially detectable through the photosynthetic red edge if the coverage and density of photoautotrophs is high enough. We concluded our discussion by sketching rubrics which might enable the behavior of NPP with spectral type and temperature to be gauged by future observations.

\section*{Data Availability Statement}
No new data were generated or analysed in support of this research.

\section*{Acknowledgements}
We thank the reviewer for the helpful and constructive report, which helped improve the quality of the paper. This research was supported in part by the Breakthrough Prize Foundation, Harvard University's Faculty of Arts and Sciences, and the Institute for Theory and Computation (ITC) at Harvard University.

%\bibliographystyle{abbrvnat}
%\bibliography{AqPhot}

\end{document}